\documentclass[a4paper, 11 pt]{article}
\pdfoutput = 1

\usepackage{jheppub, bm}
\pdfoutput=1
 \allowdisplaybreaks
\usepackage{graphicx}
\usepackage[dvipsnames]{xcolor}
\usepackage{amsmath}
\allowdisplaybreaks
\usepackage{amssymb}

\usepackage{slashed}
\usepackage[utf8]{inputenc}
\usepackage{color}
\usepackage[normalem]{ulem}
\usepackage{soul}
\usepackage{units}
\usepackage{rotating}
\usepackage{hhline,multirow,tabularx}
\usepackage{bm}
\usepackage{hyperref}
\usepackage{comment}
\usepackage[export]{adjustbox}
\usepackage{mdframed}

\def\bea#1\eea{\begin{align}#1\end{align}} 

\newcommand{\bef}{\begin{figure}[htb]\centering}
\newcommand{\eef}{\end{figure}}
\usepackage{mathtools}

\newcommand{\nn}{\nonumber}

\def\<{\langle}
\def\>{\rangle}

\def\({\left(}
\def\[{\left[}
\def\){\right)}
\def\]{\right]}

\def\cos{\hbox{cos}}
\def\sin{\hbox{sin}}
\def\ln{\hbox{ln}}

\DeclareMathOperator{\sech}{sech}

\usepackage{lipsum}
\begin{document}
\preprint{LA-UR-23-30764}

\makeatletter
\def\@fpheader{~}
\makeatother

\title{Precision three-dimensional imaging of nuclei using recoil-free jets}

\author[a]{Shen Fang,}

\author[b]{Weiyao Ke,}

\author[a,c,d]{Ding Yu Shao,}

\author[e]{and John Terry}

\affiliation[a]{Department of Physics and Center for Field Theory and Particle Physics, Fudan University, Shanghai 200438, China}

\affiliation[b]{Institute of Particle Physics and Key Laboratory of Quark and Lepton Physics (MOE), Central China Normal University, Wuhan, Hubei, 430079, China}

\affiliation[c]{Key Laboratory of Nuclear Physics and Ion-beam Application (MOE), Fudan University, Shanghai 200438, China}

\affiliation[d]{Shanghai Research Center for Theoretical Nuclear Physics, NSFC and Fudan University, Shanghai 200438, China}

\affiliation[e]{Theoretical Division, Los Alamos National Laboratory, Los Alamos, NM 87545, USA}

\emailAdd{sfang23@m.fudan.edu.cn}
\emailAdd{weiyaoke@ccnu.edu.cn}
\emailAdd{dingyu.shao@cern.ch}
\emailAdd{jdterry@lanl.gov}


\abstract{In this study, we explore the azimuthal angle decorrelation of lepton-jet pairs in e-p and e-A collisions as a means for precision measurements of the three-dimensional structure of bound and free nucleons. Utilizing soft-collinear effective theory, we perform the first-ever resummation of this process in e-p collisions at NNLL accuracy using a recoil-free jet axis. Our results are validated against Pythia simulations. In e-A collisions, we address the complex interplay between three characteristic length scales: the medium length $L$, the mean free path of the energetic parton in the medium $\lambda$, and the hadronization length $L_h$. We demonstrate that in the thin-dilute limit, where $L \ll L_h$ and $L \sim \lambda$, this process can serve as a robust probe of the three-dimensional structure for bound nucleons. We conclude by offering predictions for future experiments at the Electron-Ion Collider within this limit.}

\maketitle

\section{Introduction}
\label{sec:intro}

The forthcoming Electron-Ion Collider (EIC) is expected to serve as an instrumental apparatus for advancing our comprehension of femtoscale nucleon structures \cite{Accardi:2012qut,AbdulKhalek:2021gbh,Anderle:2021wcy,Abir:2023fpo}. It is designed to offer an unparalleled platform for the systematic investigation of the atomic nucleus, focusing specifically on its quark and gluon constituents.

Historically, the semi-inclusive deep inelastic scattering (SIDIS) process \cite{Ji:2004wu,Ji:2004xq} has served as the primary window for viewing the three-dimensional structure of matter in e-p collisions. In SIDIS, the momentum of the initial-state quark is reconstructed by correlating it with that of an observed final-state hadron. The factorization for SIDIS involves a convolution of non-perturbative inputs from the transverse-momentum-dependent (TMD) parton distribution functions (PDFs) and the TMD fragmentation functions (FFs) \cite{Collins:2011zzd,Boussarie:2023izj}.

Recent investigations at both the RHIC and the LHC have validated jets as effective tools for probing the inner structure of the nucleon \cite{Kang:2020xyq,Gao:2023ulg,STAR:2007yqh,STAR:2017akg,Aschenauer:2016our,Boer:2003tx,Vogelsang:2005cs,Bomhof:2007su,Kang:2020xez,Chien:2019gyf,Qiu:2007ey,Kang:2019ahe}. The imminent arrival of the EIC, with its high luminosity and polarized beams, promises to fully exploit the capabilities of jets in providing unique insights into the nucleon's structure. As such, it's no surprise that jet physics at the EIC is rapidly growing as a research area \cite{Liu:2018trl,Gutierrez-Reyes:2019vbx,Liu:2020dct,Makris:2020ltr,Kang:2020fka,Arratia:2020azl,Arratia:2020ssx,Kang:2020xgk,Arratia:2020nxw,delCastillo:2020omr,Kang:2021kpt,Zhang:2021tcc,Kang:2021ffh,Hatta:2021jcd,Cirigliano:2021img,Kang:2021ryr,delCastillo:2021znl,Liu:2021lan,Li:2021uww,Yan:2021htf,Yan:2022npz,Arratia:2022oxd,Burkert:2022hjz,Lee:2022kdn,Lai:2022aly,delCastillo:2023rng,Caucal:2023fsf,Caucal:2023nci}.

In past deep inelastic scattering (DIS) experiments, scientists predominantly focused on jet behaviors in the Breit frame—the frame of the virtual photon and the nucleon. Recently, there has been a surge of interest in studying observables in the lab frame of the incoming lepton and nucleon, as exemplified in hadron production \cite{Gao:2022bzi}, event shape \cite{Kang:2013nha,Li:2020bub}, and jet production \cite{Liu:2018trl}. In this frame, the transverse momentum imbalance between the final-state lepton and jet directly reflects the transverse momentum of the incoming quark. By measuring this imbalance, or the azimuthal angle decorrelation between the lepton and the jet, we obtain a more direct assessment of the TMD PDF than from SIDIS, for instance. Recent measurements at HERA \cite{H1:2021wkz} indicate that the momentum imbalance between the lepton and the jet in e-p collisions aligns well with existing theories grounded in TMD factorization. These results pave the way for future jet studies at the upcoming EIC.

While the lepton-jet correlation of \cite{Liu:2018trl,Liu:2020dct}, where a standard jet axis is used, is less sensitive to non-perturbative input, there is a trade off. SIDIS is a global observable, while the lepton-jet correlation in \cite{Liu:2018trl,Liu:2020dct} is non-global \cite{Dasgupta:2001sh}. The non-global nature of this observable is a consequence of the asymmetric treatment of soft emissions. Namely, while soft radiation exiting the jet causes the jet to recoil and thus contributes to the observed lepton-jet imbalance, soft emissions within the jet do not contribute to the observable when using a standard jet axis. The establishment of a resummation formalism for this correlation requires the resummation of non-global logarithms (NGLs). Due to this complication, resummation for this process has only been achieved at Next-to-Leading Logarithmic (NLL) accuracy. This perturbative accuracy contrasts against that of SIDIS, where the perturbative ingredients are known up to N$^3$LO+N$^4$LL accuracy (with the exception of the five loop cusp anomalous dimension), see for instance \cite{Korchemsky:1987wg, Moch:2004pa,Moch:2005id,Moch:2005tm,Idilbi:2005ni,Idilbi:2006dg,Becher:2006mr,Almelid:2015jia,Almelid:2017qju,Das:2019btv,Luo:2019szz,vonManteuffel:2020vjv,Duhr:2020seh,Ebert:2020yqt,Chen:2021rft,Duhr:2022yyp,Moult:2022xzt,Lee:2022nhh,Luo:2020epw,Herzog:2018kwj,Herzog:2017ohr}, while the non-perturbative structures of the collinear PDF and FF have been extracted from data at N$^2$LO and NLO accuracy. Thus global analyses of the TMD PDF can be achieved in full consistency at NLO+N$^2$LL accuracy from SIDIS data (where the non-perturbative contribution of the collinear FF serves as the bottleneck, see for instance \cite{Abele:2021nyo,Borsa:2022vvp,AbdulKhalek:2022laj,Bacchetta:2022awv} for recent approximate treatments of global analyses at N$^2$LO). In contrast, global analyses of the TMD PDF have only been achieved at LO+NLL accuracy of the lepton-jet correlation (where the resummation of NGLs beyond NLL significantly complicates the resummation structure\footnote{Some recent progress in NGLs resummation beyond NLL in large-$N_c$ limit can be found in \cite{Banfi:2021owj,Banfi:2021xzn,Becher:2023vrh}.}).

In this paper, we improve the perturbative accuracy of the lepton-jet correlation by using a recoiling-free jet axis \cite{Bertolini:2013iqa,Larkoski:2014uqa} instead of a standard jet axis. As the direction of the recoiling-free jet axis is insensitive to soft emissions, the use of this jet axis removes the non-global logarithms associated with the asymmetric treatment of soft radiation in the azimuthal angular distribution in the back-to-back limit \cite{Banfi:2008qs,Chien:2020hzh,Chien:2022wiq} and thus avoids the complications associated with resumming non-global logarithms beyond NLL. In this paper, we use the machinery of soft-collinear effective theory (SCET) to derive a factorization theorem tailored for back-to-back lepton-jet configurations in electron-proton (e-p) collisions. As a result, we present the first all-order resummation at NNLL accuracy for lepton-jet azimuthal distributions. Furthermore, we remark that all perturbative and non-perturbative ingredients for this process are known up to N$^2$LO+N$^3$LL accuracy, with the exception of the constant $j^{[2]}$ in the perturbative expansion for the recoil-free jet function. Nevertheless, this constant was extracted numerically \cite{Gutierrez-Reyes:2019vbx} from the Event2 generator \cite{Catani:1996vz}\footnote{A preliminary numerical results of $j^{[2]}$ are also presented in \cite{scet2023}.}. Thus the azimuthal angle decorrelation that we discuss in this paper offers the advantage of being insensitive to the non-perturbative contributions of the FF, which allows one to exceed the perturbative accuracy of SIDIS.

We further discuss that in addition to measuring vacuum TMDs, this process is particularly useful for measuring nuclear-TMDs (nTMDs) \cite{Alrashed:2021csd}. In nuclear collisions, it is well-known that the intrinsic non-perturbative parton distribution is different from the independent superposition of free nucleons \cite{EuropeanMuon:1983wih,Alrashed:2021csd}, which is often extracted by comparing ratios of heavy nuclei cross sections to those of protons or neutrons \cite{Seely:2009gt,CLAS:2019vsb,NewMuon:1995cua,NewMuon:1995tgs,NuSea:1999egr,Arnold:1983mw,NewMuon:1996yuf,NewMuon:1996gam,EuropeanMuon:1992pyr,Alde:1990im,NA3:1981yaj}. In addition to modifying the non-perturbative partonic distributions, dynamical effects, i.e., process-dependent parton/jet-nucleus interactions can also contribute to the observed difference. Developing a variety of probes in e-A within the factorization framework is essential to the disentanglement of the two effects because different observables share a set of universal nuclear PDFs, but their response to dynamical medium effects can be different. This list of observables may include SIDIS \cite{Ji:2004wu,Ji:2004xq}, jet production \cite{Li:2020rqj}, as well as the lepton-jet correlation observable discussed in this work. Because of the proliferation of energy scales in a nucleus,  factorization in e-A is fairly complicated. In this work, we will argue that in the limit of asymptotically large jet energy as measured from the nuclear rest frame, the medium effect reduces to the transverse momentum broadening of the jet axis with the jet energy loss effect being power suppressed. Additionally, we emphasize that in nuclear collisions the use of a recoil-free jet is even more useful, as it suppresses background from underlying events. 

This paper is organized as follows. In Sec.
\ref{sec:fac}, we will present a detailed study on the TMD factorization for the lepton-jet correlation in e-p and e-A collisions. In Sec. \ref{sec:Numerics}, we perform phenomenological studies for the relevant kinematics at the EIC.  Finally, we summarize our paper in Sec. \ref{sec:Sum}.

\section{Factorization and resummation}\label{sec:fac}

\subsection{Kinematics}

\begin{figure}
    \centering
    \includegraphics[scale = 0.65]{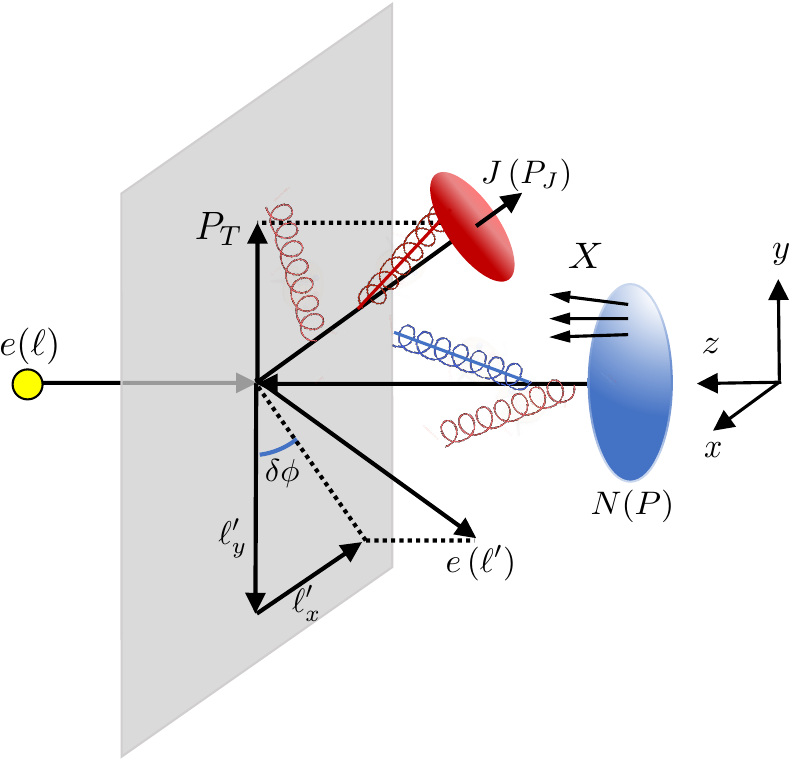}
    \caption{The azimuthal angular decorrelation for the recoil-free jet. The blue, red, and pink gluons represent collinear, jet, and soft radiation, respectively.}
    \label{fig:kinematics}
\end{figure}

We denote the momenta in the azimuthal angle decorrelation of the lepton-jet in e-N collisions as
\begin{align}\label{eq:pro}
    e(\ell) + N(P) \rightarrow e(\ell') + J(P_J) + X\,,
\end{align}
where $e(\ell)$ denotes the incoming electron with momentum $\ell$, $N(P)$ symbolizes a nucleon or nucleus with momentum $P$, and $e(\ell')$ refers to the scattered electron with momentum $\ell'$. The final state also includes a jet $J$ with momentum $P_J$ and unobserved particles which are collectively represented as $X$. In the center-of-mass (CM) frame of the incident lepton and hadron, the four-momenta of the initial-state lepton and hadron can be expressed in terms of the lepton-proton CM energy as:
\begin{align}
    P^\mu = \sqrt{S}\, \frac{n^\mu}{2}\,,
    \qquad
    \ell^\mu = \sqrt{S}\, \frac{\bar{n}^\mu}{2}\,,
\end{align}
where $n^\mu$ and $\bar{n}^\mu$ are the light-cone coordinates and we use the conventions $n^\mu=(1,0,0,1)$ and $\bar{n}^\mu=(1,0,0,-1)$ in Cartesian space-time coordinates. In taking this parameterization, the $z$ direction is defined along the nucleon beam while the electron beam is in the negative $z$ direction. The final-state leptonic momentum is parameterized most conveniently in terms of the event inelasticity $y = 1-P\cdot \ell'/P\cdot l$ and the lepton's transverse momentum as 
\begin{align}
    {\ell'}^\mu = \sqrt{S}\left\{1-y, \frac{{\ell'_T}^2}{S(1-y)}, \frac{\ell'_x}{\sqrt{S}}, \frac{\ell'_y}{\sqrt{S}}\right\}\,.
\end{align}
Here we have organized the contributions using light-cone coordinates as $\left\{n\cdot \ell',\bar{n}\cdot \ell',\bm{\ell}'_T\right\}$, where the curly brackets denote that we are working in light-cone coordinates. Further, we represent $\ell'_T = |\bm{\ell}'_T|$ as the magnitude of the transverse momentum of the final-state lepton, while $\delta \phi$ represents the azimuthal angle which characterizes the momentum of the lepton out of the $y$-$z$ plane. This angle and the kinematics of the process are depicted in Fig.~\ref{fig:kinematics}. 

By momentum conservation, the four momentum of the intermediate photon is given by $q = \ell-\ell'$. From this parameterization, we can define the kinematic variables as
\begin{align}
    & Q^2 = -(\ell-\ell')^2 = \frac{{\ell'_T}^2}{1-y
    }\,, 
    \qquad
    x_B = \frac{Q^2}{2 P\cdot q} = \frac{Q^2}{y S}\,,
    \\
    & E_J^{\rm rest} = \frac{Q^2}{2 M x_B} = \frac{y S}{2 M}\,,
    \qquad
    \hat{t} = -Q^2\,, 
    \qquad 
    \hat{s} = x_B S\,, \nn
\end{align}
where in the second line, $E_J^{\rm rest}$ represents the energy of the jet in the rest frame of the nucleus while $\hat{u} = -\hat{s}-\hat{t}$ is the final partonic Mandelstam variable, and $M$ is the mass of the free or bound proton.

Lastly, the momenta of the final-state jet in the CM frame can be parameterized by $P_J^\mu = E_J\left(1,\hat{r}\right)$ where $\hat{r}$ is a radial unit vector with azimuthal and polar angles $\phi_J$, and $\theta_J$ while $E_J$ denotes the energy of the jet in the CM frame. The polar angle is related to the rapidity of the jet through the relation $\theta_J = \arcsin\left(\sech(y_J)\right)$. In this study, we exploit the rotation invariance of the initial state to take the transverse momentum of the jet to be in the $y$ direction such that the jet has a transverse momentum given by $\bm{P}_{T} = E_J\,\sin{\theta_J}\,\hat{y}$. By parameterizing the jet this way, we can define the jet light-cone coordinates
\begin{align}
    n_J^\mu        = \left(1, \hat{r}\right)\,,
    \qquad
    \bar{n}_J^\mu  = \left(1,-\hat{r}\right)\,.
\end{align}
These light-cone coordinates are constructed such that $n_J\cdot P_J = 0$, while $n_J\cdot \bar{n}_J = 2$.

\subsection{Factorization in e-p}
The transverse momentum imbalance of the lepton-jet pair is defined as $\bm{q}_T = \bm{\ell}_T'+\bm{P}_T$. By exploiting the rotational invariance to place $\bm{P}_T\sim \hat{y}$, the component $q_x = \ell'_T \delta\phi$ becomes correlated with the transverse momenta of the partons. In the back-to-back limit where $\delta\phi \ll 1$, measurements of this component become sensitive to the intrinsic momentum of the initial state quark. In this limit, emissions from the soft, collinear, and jet sectors will contribute to the $\delta\phi$ distribution so that the QCD modes that contribute to the cross section are given by
\begin{align}
  \textbf{ hard}:&~~p_h^\mu \sim \ell'_T (1,1,1), \nn\\[1pt]
  {\color{black} n \textbf{-collinear}}:&~~p_c^\mu\sim \ell'_T\,(\delta\phi^2,1,\delta\phi), \nn\\[1pt]
  {\color{black} \textbf{soft}}:&~~p_s^\mu\sim \ell'_T\,(\delta\phi,\delta\phi,\delta\phi), \nn\\[1pt]
  { \color{black} n_{J} \textbf{-collinear}}:&~~p_J^\mu\sim \ell'_T\,(\delta\phi^2,1,\delta\phi)_J\,, \nn  
\end{align}
where the $J$ subscript on the jet momentum scaling represents that the $p_J$ momentum is decomposed using the jet light-cone components. In this mode analysis, we have also introduced the hard modes that are used to match QCD onto SCET. 

Each mode presented above will contain a natural renormalization scale $\mu$, while the collinear, soft, and jet modes will all depend on a rapidity scale $\nu$. Namely for this case, the momentum characterizing the jet axis contains transverse momentum relative to the total momentum of the jet, which introduces a dependence on the rapidity scale. The size of the scales associated with each mode are given schematically as
\begin{align}\label{eq:scale-can}
    & \mu_H \sim \nu_\mathcal{J} \sim \nu_B \sim \ell'_T\,,
    \\
    & \mu_\mathcal{J} \sim \mu_B \sim \mu_S \sim \nu_S \sim \ell'_T \delta \phi \,,
\end{align}
where $\mu_i$ and $\nu_i$ denote the renormalization and rapidity renormalization scales of the $i$ mode, respectively. Taking into account these modes and using standard methods of factorization theorems in SCET, the factorized cross section is given by
\begin{align}
    \frac{d\sigma}{d^2\ell'_{T}\, dy\, dq_x} = \frac{\sigma_0}{1-y} & H\left(Q,\mu\right)\mathcal{C}\left[B\,\mathcal{J}\,S\right]\,,
\end{align}
where the Born cross section is given by
\begin{align}
    \sigma_0 = \frac{\alpha_{\rm em}^2}{
     S Q^2}\frac{2\left(\hat{s}^2+\hat{u}^2\right)}{\hat{t}^2}\,,
\end{align}
while $H$, $S$, $B$, and $\mathcal{J}$ are the hard, soft, beam, and jet functions. To simplify this expression, we have used the short-hand notation
\begin{align}\label{eq:conv}
\mathcal{C}\left[B\,\mathcal{J}\,S\right] & = \sum_q e_q^2 \int dk_x\, d\lambda_x\, d p_x\, \delta\left(q_x-k_x-p_{x}-\lambda_x\right) \nn \\
    &\times B_{q/N}\left(x_B,k_x,\mu,\zeta_B/\nu^2\right) \mathcal{J}_{q}\left(p_x,\mu,\zeta_\mathcal{J}/\nu^2\right)  S\left(\lambda_x,n\cdot n_J,\mu,\nu\right) \,,
\end{align}
where $\zeta_{B/\mathcal{J}}$ denote the Collins-Soper parameters of the beam and jet functions which are explicitly given by
\begin{align}
    \zeta_B = \left(\bar{n}\cdot k\right)^2\,,
    \qquad
    \zeta_\mathcal{J} = \left(\bar{n}_J\cdot P_J\right)^2\,.
\end{align}
We see in Eq.~\eqref{eq:conv} that the soft function depends on $n\cdot n_J$, which differs from the back-to-back soft function in SIDIS for instance. Lastly, we use $k_x$, $p_x$, and $\lambda_x$ to denote the transverse momentum of the incoming quark relative to the parent hadron, the transverse momentum of the jet axis relative to the jet momentum, and the transverse momentum generated by the soft emissions. After going to Fourier space, the convolution is simply given by
\begin{align}
\mathcal{C}\left[B\,\mathcal{J}\,S\right] & = \sum_q e_q^2 \int \frac{db}{2\pi}\cos{(b\,q_x)} B_{q/N}\left(x_B,b,\mu,\zeta_B/\nu^2\right)  \mathcal{J}_{q}\left(b,\mu,\zeta_\mathcal{J}/\nu^2\right)\, S\left(b,n\cdot n_J,\mu,\nu\right) \,.
\end{align}

\subsection{Resummation in e-p}
To generate perturbative predictions in the back-to-back region, we need to perform resummation of the large logarithms present in the perturbative expansion of each mode within the factorization. In this section, we perform this resummation to NNLL accuracy. 

First of all, the hard function satisfies the RG equation
\begin{align}
    \frac{d}{d \ln{\mu}}H\left(Q,\mu\right) = \gamma^H_\mu(\mu) H\left(Q,\mu\right)\,,
\end{align}
and the solution to this equation is given by
\begin{align}
    H\left(Q,\mu\right) = H\left(Q,\mu_H\right)U_H\left(\mu_H,\mu\right)\,,
\end{align}
where the logarithms are resumed in the Sudakov exponential
\begin{align}
    U_H(\mu_H,\mu)=\exp\left[\int_{\mu_H}^\mu \frac{d\mu'}{\mu'} \gamma^H_\mu\left(\mu'\right) \right]\,,
\end{align}
and the anomalous dimension for this process is given up to NNLL by the expression
\begin{align}\label{eq:anom-hard}
    \gamma^H_\mu(\mu) = 2C_F\gamma_{\rm cusp}(\alpha_s) \ln\left(\frac{Q^2}{\mu^2}\right)+4\gamma_q(\alpha_s)\,.
\end{align}
In the expression for the hard anomalous dimension, $\gamma_{\rm cusp}$ and $\gamma_q$ are the cusp and non-cusp anomalous dimension which is given up to NNLL in the Appendix \ref{sec:anom}.

The beam, soft, and recoil-free jet functions depend on two scales, thus their evolution is governed by the coupled evolution equations
\begin{align}
    \frac{d}{d \ln{\mu}}F\left(b,\mu,\nu\right) &= \gamma^F_\mu(\mu,...) F\left(b,\mu,\nu\right)\,,
    \\
    \frac{d}{d \ln{\nu}}F\left(b,\mu,\nu\right) &= \gamma^F_\nu(b,\mu) F\left(b,\mu,\nu\right)\,,
\end{align}
where $F \in \{B,\mathcal{J},S\}$ and the ellipsis represents an arbitrary dependence on additional scales. The soft anomalous and rapidity anomalous dimensions are given up to NNLL by
\begin{align}\label{eq:anom-soft}
    \gamma^S_\mu\left(\mu,\frac{\mu}{\nu}\right) & = -2 C_F\gamma_{\rm cusp}(\alpha_s) \ln\left(\frac{\nu^2\, n\cdot n_J}{2\mu^2}\right) - 2 C_F \gamma_s(\alpha_s)\,, \nn \\
    \gamma^S_{\nu}\left(b,\mu\right) & = 2C_F\left[\int_{\mu^2}^{\mu_b^2} \frac{d\bar{\mu}^2}{\bar{\mu}^2}\gamma_{\rm cusp}(\alpha_s)+\gamma_r(\alpha_s)\right]\,,
\end{align}
where $\gamma_r$ is the rapidity anomalous dimension. Similarly, the anomalous dimensions of the beam and jet functions are given by
\begin{align}
\gamma^{B/\mathcal{J}}_{\mu}\left(\mu,\frac{\zeta_{B/\mathcal{J}}}{\nu^2}\right) = &-C_F\gamma_{\rm cusp}(\alpha_s) \ln\left(\frac{\zeta_{B/\mathcal{J}}}{\nu^2}\right) +\gamma_{B/\mathcal{J}}(\alpha_s)\,, \nn \\
    \gamma^{B/\mathcal{J}}_{\nu}\left(b,\mu\right) = & -\frac{1}{2}\gamma^S_\nu(b,\mu)\,.
\end{align}
Due to the simplicity of the color space in e-p collisions, one can easily verify the renormalization group consistency relation for this process
\begin{align}    
\gamma^H_\mu(\mu)+ \gamma^S_\mu\left(\mu,\frac{\mu}{\nu}\right)+\gamma^B_\mu\left(\mu,\frac{\zeta_B}{\nu^2}\right)+\gamma^\mathcal{J}_\mu\left(\mu,\frac{\zeta_\mathcal{J}}{\nu^2}\right) &= 0\,, \nn
\\
 \gamma^S_\nu(b,\mu)+\gamma^B_\nu(b,\mu)+\gamma^\mathcal{J}_\nu(b,\mu) &= 0\,.
\end{align}
From the above expression, we can define properly subtracted TMDs which do not depend on the rapidity scale through the subtraction
\begin{align}
f_{q/N}\left(x_B,b,\mu,\zeta_f\right) & =  B_{q/N}\left(x_B,b,\mu,\zeta_B/\nu^2\right)  \sqrt{S_{n \bar{n}}\left(b,\mu,\nu\right)}\,, \\
J_q\left(b,\mu,\zeta_J\right) &= \mathcal{J}_q\left(b,\mu,\zeta_\mathcal{J}/\nu^2\right)\, \frac{S\left(b,n\cdot n_J,\mu,\nu\right)}{\sqrt{S_{n \bar{n}}\left(b,\mu,\nu\right)}}\,,
\end{align}
where we have introduced the Collins-Soper scales for the subtracted functions as 
\begin{align}
    \sqrt{\zeta_f} = \sqrt{\zeta_B}\,,
    \qquad
    \sqrt{\zeta_J} = \frac{n\cdot n_J}{2}\sqrt{\zeta_\mathcal{J}}\,,
\end{align}
and we note that the factor of $n\cdot n_J/2$ enters from the soft factor. In performing this soft subtraction, we have introduced the back-to-back soft function which is given at one loop in Eq.~\eqref{eq:loop-softnn} while the anomalous dimensions of this function are given at NNLL in Eqs.~\eqref{eq:anom-soft-nn}. This soft subtraction removes the dependence of the soft subtracted TMD PDF and the jet function on the scale $\nu$. After taking into consideration the soft subtraction, we arrive at the expected relation $\zeta_J \, \zeta_f = Q^4$. 

At NNLL accuracy, the fixed order expressions for the hard, soft, beam, and jet functions should all be carried out to NLO accuracy. The TMD beam function can be perturbatively matched onto the collinear PDFs via an operator product expansion
\begin{align}
    f_{q/N}\left(x_B,b,\mu,\zeta_f\right) & = \sum_i \int_x^1 \frac{d\hat{x}}{\hat{x}} C_{q/i}\left(\frac{x_B}{\hat{x}},b,\mu,\zeta_f\right)  f_{i/N}\left(\hat{x},\mu\right) \nn 
    \\
    & = \left[C\otimes f\right]_{q/N}\left(x_B,b,\mu,\zeta_f\right)\,,
\end{align}
where $f_{i/N}$ in the convolution denotes the collinear PDF and we note that this matching holds only in the small $b$ region, while at large $b$ non-perturbative contribution can enter. After performing the soft subtraction, the explicit one-loop expressions for these functions are given in Eq.~\eqref{eq:loop-coll-sub-qq}. We see in these expressions that large logarithms which go like $\ln\left(\zeta/\mu\right)$ enter into the perturbative expressions for the beam and jet functions, which need to be resumed. This resummation can be performed by introducing initial scales $\sqrt{\zeta_{fi}}$ and $\sqrt{\zeta_{Ji}}$ and evolving these distributions to the Collins-Soper scales by solving the Collins-Soper evolution equations. Therefore, we have
\begin{align}
    \frac{d}{d\ln\mu}  F(b,\mu,\zeta) & = \gamma_\mu^F(\mu,\zeta)F(b,\mu,\zeta)\,,\\
    \frac{d}{d\ln\zeta}F(b,\mu,\zeta) & = \frac{1}{2}\gamma_\zeta^F(\mu,b)F(b,\mu,\zeta)\,,
\end{align}
for $F\in\left\{f,J\right\}$ and the anomalous dimensions are given by
\begin{align}
    \gamma_\mu^{f}(\mu,\zeta) & = \gamma_\mu^{B}\left(\mu,\frac{\zeta_{B}}{\nu^2}\right)+\frac{1}{2}\gamma_\mu^{S_{n \bar{n}}}\left(\mu,\frac{\mu}{\nu}\right)\,,\nn\\
    \gamma_\mu^{J}(\mu,\zeta) & = \gamma_\mu^{\mathcal{J}}\left(\mu,\frac{\zeta_{\mathcal{J}}}{\nu^2}\right)+\gamma_\mu^{S}\left(\mu,\frac{\mu}{\nu}\right)-\frac{1}{2}\gamma_\mu^{S_{n \bar{n}}}\left(\mu,\frac{\mu}{\nu}\right)\,,\nn\\
    \gamma_\zeta^{f/J}(\mu,b) & = -\gamma_\nu^{B/\mathcal{J}}(b,\mu)\,. 
\end{align}
The solutions to the evolution equations are given by 
\begin{align}
    F(b,\mu,\zeta) = F(b,\mu_F,\zeta_F) U_F\left(\mu_F,\mu;\zeta\right)Z_F\left(\zeta_F,\zeta;\mu_F\right)\,,
 \end{align}
where the resummation of logarithmic terms is performed in the Sudakovs
\begin{align}
    U_F\left(\mu_F,\mu;\zeta\right) & = \exp\left[\int_{\mu_F}^\mu \frac{d\mu'}{\mu'}\gamma^F_\mu\left(\mu',\zeta\right)\right]\,, \\
    Z_F\left(\zeta_F,\zeta;\mu\right) & = \exp\left[\int_{\zeta_F}^\zeta \frac{d\zeta'}{\zeta'}\frac{1}{2}\gamma^F_\zeta\left(b,\mu\right)\right]\,.
\end{align}
By studying the one-loop perturbative expressions, we arrive at the natural scale for the hard, TMD PDF, and jet functions 
\begin{align}\label{eq:scale-can-exp-zeta}
     \mu_H = Q\,, \quad\quad
     \mu_f = \mu_J = \sqrt{\zeta_{fi}} = \sqrt{\zeta_{J i}} = \mu_b =2 e^{-\gamma_E}/b\,.
\end{align}
With these scale choices,the final expression for the factorized cross section can be expressed as
\begin{align}\label{eq:res}
    & \frac{d\sigma}{d^2\ell'_{T}\, dy\, d\delta\phi} = \frac{\sigma_0\, \ell'_T}{1-y} H\left(Q,\mu\right)\int_0^\infty \frac{db}{\pi} \cos\left(b \ell'_T \delta\phi\right) 
    \sum_q e_q^2\, f_{q/N}\left(x_B,b,\mu,\zeta_f\right) J_q\left(b,\mu,\zeta_J\right)  \,,
\end{align}
where TMD PDFs and jet functions are given by 
\begin{align}\label{eq:TMDPDF}
    f_{q/N}  \left(x_B,b,\mu,\zeta_f\right)  & = \left[C\otimes f\right]_{q/N}\left(x_B,b,\mu_f,\zeta_{fi}\right) \, U_{\rm NP}^{f}(x_B,b,A,Q_0,\zeta_f) \nn \\
&\times \exp\left[\int_{\mu_f}^{\mu} \frac{d\mu'}{\mu'}\gamma^f_\mu\left(\mu',\zeta_f\right)\right]
\left(\frac{\zeta_f}{\zeta_{fi}}\right)^{\frac{1}{2}\gamma^f_\zeta\left(b,\mu_f\right)}\,,
\end{align}
and
\begin{align}\label{eq:TMDjet}
J_{q}  \left(b,\mu,\zeta_J\right) &= J_{q}\left(b,\mu_J,\zeta_{Ji}\right) \, U_{\rm NP}^{J}(b,A,Q_0,\zeta_J)\nn \\
&\times \exp\left[\int_{\mu_J}^{\mu} \frac{d\mu'}{\mu'}\gamma^J_\mu\left(\mu',\zeta_J\right)\right]
\left(\frac{\zeta_J}{\zeta_{Ji}}\right)^{\frac{1}{2}\gamma^J_\zeta\left(b,\mu_J\right)} \,,
\end{align}
respectively. In these expressions, we have introduced the non-perturbative Sudakov terms $U_{\rm NP}$. We note that while the jet functions are often taken to be purely perturbative, the recoil-free jet will contain non-perturbative corrections due to the initial scale $\zeta_{Ji} = \mu_b^2$. Namely, at large $b$, the Collins-Soper kernel becomes non-perturbative and to systematically describe this process, we need to take this contribution into account. We note, however, that the rapidity anomalous dimension of the jet is identical to that of the TMDs. As a result, while this jet function contains this non-perturbative contribution, the form of this function can be taken from global analyses of TMDs. Lastly, we note that for phenomenology, we will take the scale choice that $\mu = Q$ while we note that the cross section depends on the product $\zeta_f \zeta_J$ instead of either of the Collins-Soper parameters. As a result, we can always take $\sqrt{\zeta_J}=\sqrt{\zeta_f}=Q$ for numerical calculations. 

\subsection{Factorization and resummation in e-A}

The cold nuclear medium introduces additional scales to the system that complicate the derivation of the factorization and resummation formalism, namely the scales associated with the finite size of the medium, the mean-free path of partons, and the hadronization length of the energetic mode passing through the medium. We will denote these length scales $L\sim A^{1/3}/\Lambda$, $\lambda\sim 1/\Lambda$, and $L_h\sim E_J^{\rm rest}/\Lambda^2$, where $\Lambda$ is the QCD non-perturbative scale. For simplicity, all these scales are understood and compared in the rest frame of the nucleus.  Working in the CM frame, these scales become $L\sim A^{1/3}/\sqrt{S}$, $\lambda\sim 1/\sqrt{S}$, and $L_h\sim E_J/\Lambda^2$. Even though the factorization in the medium is a complicated problem that is still under intensive study, here we seek limits where the leading medium effects only appear as a broadening factor, while other forms of medium corrections are power suppressed.

In the region where $L_h < L$, known as the large-medium limit, the partonic constituents of the jet will hadronize within the nuclear medium, introducing non-perturbative contributions associated with in-medium fragmentation and hadronic collisions. In the limit that $L_h \gg L$, however, the situation is drastically simplified. In this region, the medium is considered `thin' and these non-perturbative contributions become power suppressed by the small parameter $v = L/L_h \sim 2 M x_B A^{1/3}/Q^2 \ll 1$, where $M$ is the mass of the struck nucleus. Nevertheless, final-state QCD interactions must be considered, and the partonic constituents of the jet mode undergo forward scattering with the medium via Glauber interactions. Such Glauber interactions are commonly modeled using a screened Coulomb potential \cite{Gyulassy:1993hr,Gyulassy:2002yv}, 
\begin{align}
    \frac{d\sigma_G}{d^2 {\bf q}_T} = \frac{\alpha_{s, \rm eff}}{\pi}\frac{1}{({\bf q}_T^2+\xi^2)^2}\,,
\end{align}
where $\xi$ is a non-perturbative mass scale that cuts the partonic interaction in the infrared, ${\bf q}_T$ is the transverse momentum of the Glauber mode, and $\alpha_{s, \rm eff}$ is the coupling between the Glauber gluon and the color sources in the medium. At large $q$, one can expect that it is the perturbative QCD coupling. However, in the medium, one is also sensitive to the physics near $\xi^2$, thus the subscript ``eff''. The average number of Glauber collisions between the jet and the medium is called the opacity parameter $\chi=L/\lambda$. These forward interactions deflect the direction of the jet, leading to transverse momentum broadening and modifications to the jet function \cite{Gyulassy:2002yv,Kang:2017frl}. At the LO in the jet emissions, the momentum broadening of the jet can be obtained by summing over all orders in the opacity parameter $\chi$. The closed form for the screened Coulomb potential \cite{Gyulassy:2002yv} as
\begin{align}
    J_q^A(b,\mu,\zeta_J) = J_q(b,\mu,\zeta_J) e^{\chi [\xi b K_1(\xi b)-1]}\,,
\end{align}
were the medium modifications are accounted for in the exponential.

At the NLO, we distinguish collinear emissions whose formation time is comparable to, or larger than the medium size and soft emissions that are frequently formed inside the medium. 
Medium-induced collinear gluon emissions can carry away a significant amount of energy from the parton and modify the in-medium jet function. However, these emissions are coherent over the entire finite-size medium and lead to a destructive interference known as the QCD Landau-Pomeranchuk-Migdal (LPM) effect \cite{Gyulassy:2000er}. As a result, in the dilute limit, modifications from medium-induced collinear emissions are power suppressed by $\chi \frac{\xi^2 L}{E} \sim \chi v$. Higher-power corrections can be obtained systemically in the opacity expansion approach \cite{Gyulassy:2000er}. Alternatively, in the dense limit $\chi\gg 1$, all-order opacity resummation becomes important \cite{Baier:1994bd,Baier:2000mf}.

Medium-induced soft emissions are incoherent, frequent emissions induced by each individual scattering center in the medium. Even though such emissions cannot change the jet energy at leading power, they are not suppressed by the LPM effect either. Multiple soft emissions need to be resumed into the broadening of jet transverse momentum in the nucleus. The effect of multiple soft emissions can be absorbed into a redefinition (a rapidity RG) of the medium broadening factor, which was used to study the medium-modified jet functions in \cite{Vaidya:2021vxu}. It resums logarithm like $\ln A^{1/3}$. This has been performed either in terms of the renormalization of nuclear matter transport parameters or in the TMD framework in the Drell-Yan process in the upcoming work \cite{Ke:2024}. Therefore, momentum broadening also receives radiative correction, but here, we will only estimate this effect phenomenologically by varying the magnitude of the effective collisional opacity $\chi$.

In this paper, we are interested in addressing the question as to how lepton-jet correlations can be used to image the three-dimensional non-perturbative structure of the nucleon in the atomic nuclei. For this purpose, momentum broadening from final-state multiple scatterings represents a background. In the nuclear rest frame, the jet parton's momentum is broadened in the two-dimensional space perpendicular to its direction of motion. In general, the broadening effect transforms non-trivially under boost to a different reference frame where the medium acquires a collective velocity \cite{Sadofyev:2021ohn}. However, we note that the current observable is constructed in a way that it is only sensitive to the broadening in the $x$ direction, which is invariant when we boost along the beam axis. Therefore, we expect the expression for the broadening factor obtained in the nuclear rest frame will hold in the collider frame as well. From our previous discussion, final-state non-perturbative effects can contribute in the large medium limit. However, for jet production at large $Q$, the system is described in the thin medium limit and these non-perturbative contributions become power suppressed and the final-state interactions are partonic. In this limit, the final-state interactions can be considered as perturbative contributions which enter in the opacity expansion, which drastically simplifies the system, introducing at most a few non-perturbative cold nuclear matter transport parameters. Further, in the dilute limit, the opacity terms do not need to be resumed and thus the resummation structure of the process is left unchanged. We then arrive at the compact factorization and resummation formula for this process
\begin{align}\label{eq:res-eA}
     \frac{d\sigma_A}{d^2\ell'_{T}\, dy\, d\delta\phi}& = \frac{\sigma_0\, \ell'_T}{1-y}\, H\left(Q,\mu\right)\int_0^\infty \frac{db}{\pi} \cos\left(b \ell'_T \delta\phi\right) 
    \sum_q e_q^2\, f_{q/A}\left(x_B,b,\mu,\zeta_f\right)J_q^A\left(b,\mu,\zeta_J\right) ,
\end{align}
where all medium contributions are accounted for in the nTMD PDF and the jet function.

\section{Numerical results}\label{sec:Numerics}
\subsection{Parameterization}
By taking the canonical scale choice, we have introduced the natural scale $\mu_b = 2 e^{-\gamma_E}/b$ in Eq. \eqref{eq:scale-can-exp-zeta}. However, in the integration, we encounter an issue in the region where $\mu_b \sim \Lambda_{\rm QCD}$, which is avoided using the $b_*$-prescription \cite{Collins:2014jpa,Aidala:2014hva,Sun:2014dqm,Landry:2002ix,Konychev:2005iy,Bacchetta:2017gcc,Bacchetta:2022awv}. In this work, we follow the standard $b_*$-prescription where
\begin{align}\label{eq:bstar}
  b_* \equiv b /\sqrt{1+b^2/b_{\rm max}^2}\,,
  \qquad
  \mu_{b_*} = 2 e^{-\gamma_E}/b_*\,,
\end{align}
as in \cite{Collins:1984kg}, where we choose $b_{\rm max}=1.5$ GeV$^{-1}$. To parameterize the non-perturbative Sudakov of the vacuum TMD PDF in Eq.~\eqref{eq:TMDPDF}, we use the parameterization of Refs. \cite{Sun:2014dqm,Kang:2015msa,Echevarria:2020hpy,Alrashed:2021csd}
\begin{align}
    U_{\rm NP}^f(x,b,A,Q_0,Q) = \exp\left[-g_1^A b^2-\frac{g_2}{2}  \ln \frac{Q}{Q_0} \ln \frac{b}{b_*}\right], \nn
\end{align}
and
\begin{align}
\label{eq:aN}
g_1^A = g_1^f + a_N (A^{1/3}-1)\,,
\end{align}
where the authors chose an $x$-independent parameterization,  $g_1^f=0.106\,\mathrm{GeV}^{2}$ represents the intrinsic width of a free proton, $a_N = 0.016$ GeV$^{2}$ characterizes the broadening associated with multiple scattering of the initial-state quark with the nuclear medium, while $g_2=0.84$ and $Q_0^2=2.4 \, \mathrm{GeV}^2$ characterize the non-perturbative behavior of the Collins-Soper kernel (rapidity anomalous dimension). Thus, for the jet function, we have
\begin{align}
    U_{\rm NP}^J(b,A,Q_0,Q) = \exp\left[-\frac{g_2}{2}  \ln \frac{Q}{Q_0} \ln \frac{b}{b_*}\right] \,.
\end{align}
To parameterize the vacuum collinear PDF, we use the CT14nlo parameterization \cite{Dulat:2015mca} while we parameterize the collinear dynamics of the nTMDPDF using the EPPS16 parameterization given in \cite{Eskola:2016oht}.

\begin{figure}
    \centering
    \includegraphics[scale = 0.6]{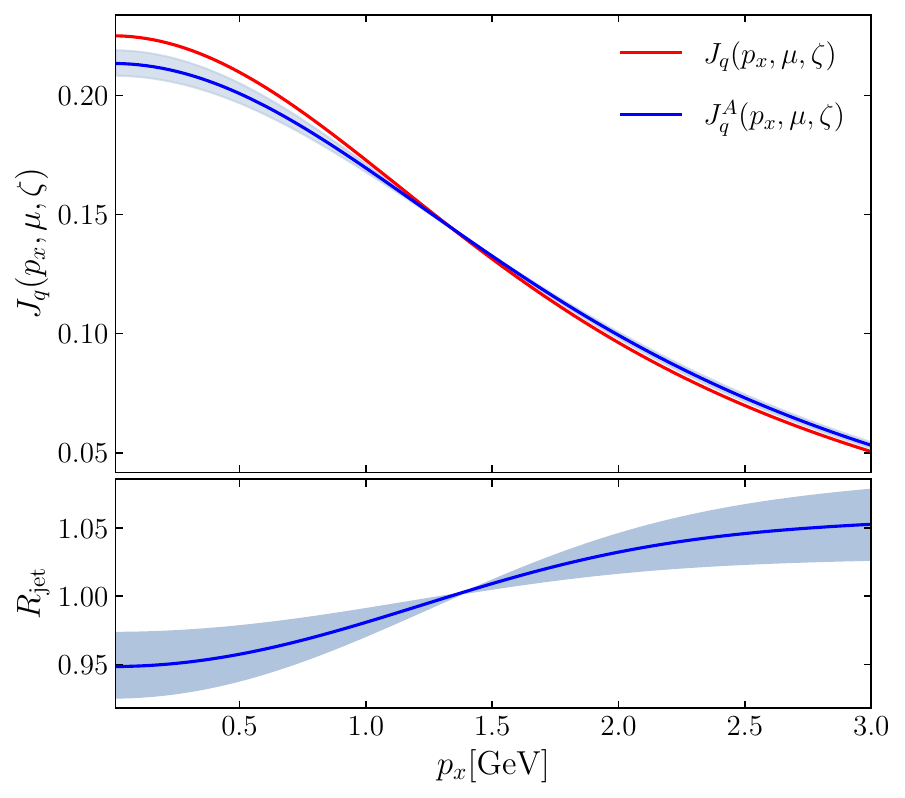}
    \caption{Red: the recoil-free jet function as a function of transverse momentum at $\mu^2 = \zeta = 100$ GeV$^2$. Blue: the recoil-free jet function with the medium modification where the blue band has been generated by varying the $\rho_G$ parameter by a factor of $1-3$. The central curve is generated at $2 \rho_G$.}
    \label{fig:jet}
\end{figure}

To parameterize the broadening effects to the jet, we first note that the opacity can be written in terms of the density of the medium $\rho_G$, the screening mass $\xi$, and the length of the medium as 
\begin{align}
    \chi = \frac{\rho_G L}{\xi^2}\alpha_s(\mu_{b_*}) C_F\,.
\end{align}
In this paper, we take the parameter values $\rho_G=0.4$~{fm}$^{-3}$, $\xi^2 = 0.12$ GeV${}^2$ from a recent comparison between SCET$_{\rm G}$ (SCET with Glauber gluon) calculation and the collinear fragmentation function in e-A from the HERMES experiment \cite{Ke:2023ixa}. The path length is $L = 3/4\times A^{1/3} \,1.2$ fm, where the $3/4$ factor is a result of geometrically averaging the location of the jet production in the nucleus.

To demonstrate the effect of the medium, we introduce the momentum space jet functions
\begin{align}\label{eq:jet-mom}
J_q^{(A)}\left(p_x,\mu,\zeta\right) &= \int \frac{db}{2\pi} \cos\left(b p_x\right) J_q^{(A)}\left(b,\mu,\zeta\right)\,.
\end{align}
The jet functions are plotted in Fig.~\ref{fig:jet} at the scale choice $\mu^2 = \zeta = 100$ GeV$^2$. The vacuum jet function is plotted in red while the modified jet function is plotted in blue. In the denominator, we plot the ratio
\begin{align}
    R_{\rm jet}(p_x,\mu,\zeta) = \frac{J_q^A(p_x,\mu,\zeta)}{J_q(p_x,\mu,\zeta)}\,.
\end{align}
In this ratio plot, we can clearly see that while the nuclear-modified jet function is suppressed at small $p_x$ and enhanced at large $p_x$. This behavior results from the exponential in the medium-modified jet function. In Ref. \cite{Ke:2024}, it was estimated that the induced radiation can enhance the broadening by about a factor of three in a heavy nucleus $A\approx 200$. Therefore, in Fig. \ref{fig:jet}, we have generated an uncertainty band by varying $\rho_G$ by a factor of 1-3 to estimate the impact of the medium-induced radiative broadening discussed for instance. To guarantee the numerical convergence of the Fourier transformation \eqref{eq:jet-mom}, we introduce a damping factor $\exp(-0.2\times b^2)$ to mitigate the oscillatory behavior inherent in the integration process.
 
\subsection{Predictions for e-p and e-A}

\begin{figure}
    \centering
    \includegraphics[scale = 0.65]{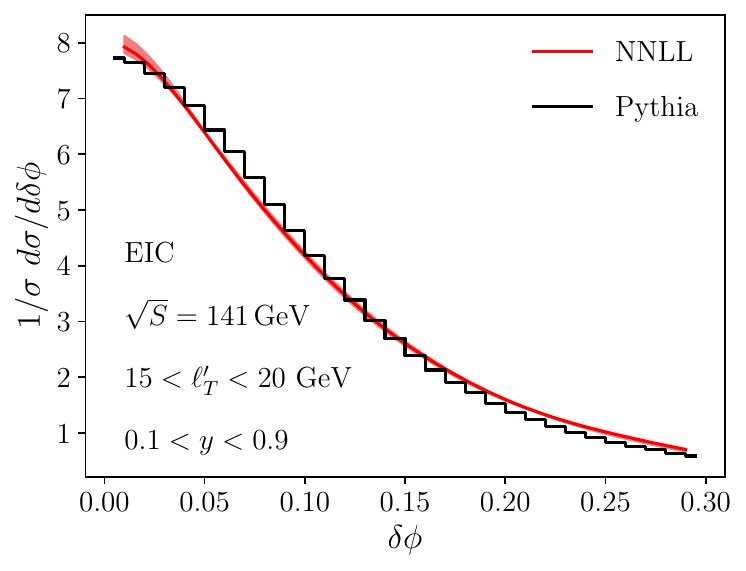}
    \caption{Normalized azimuthal angular decorrelation in e-p collisions. Integration is over lepton momentum ($15 \, \text{GeV} < \ell_T' < 20 \, \text{GeV}$) and event inelasticity ($0.1 < y < 0.9$). The black line represents Pythia simulation results, while the red line indicates predictions from the resummation formalism \eqref{eq:res}. In the predictions, the hard scale $ \mu_H $ varies between $ Q/2 $ and $ 2\,Q $, with uncertainties denoted by red bands. }
    \label{fig:resum}
\end{figure}

Due to the lack of data on lepton-jet correlations with a recoil-free jet axis, we demonstrate the fidelity of our work by comparing predictions in our theoretical formalism against a Pythia 8.3 \cite{Bierlich:2022pfr} simulation for e-p collisions at EIC kinematics. This is depicted in Fig.~\ref{fig:resum}. For this process, the energy configuration is taken from the EIC yellow paper \cite{AbdulKhalek:2021gbh} at $\sqrt{S} = 141$ GeV, and the events have been integrated over the lepton momentum and event inelasticity range of $15$ GeV $< \ell_T' < 20$ GeV and $0.1<y<0.9$. The black line in the figure corresponds to the Pythia simulation results, while the red line showcases the numerical predictions from our resummation formalism \eqref{eq:res}. In the latter, the hard scale $ \mu_H $ is varied between $ Q/2 $ and $ 2\,Q $, and the corresponding uncertainties are shown as the red bands. Obviously, it only has a mild impact in the small $\delta \phi$ limit, and the scale uncertainties can be further suppressed after including N$^3$LL resummation. Through this analysis, we see that in the region $\delta\phi < 0.3$, our formalism is strongly consistent with the Pythia simulation.

\begin{figure}
    \centering
    \includegraphics[scale = 0.6]{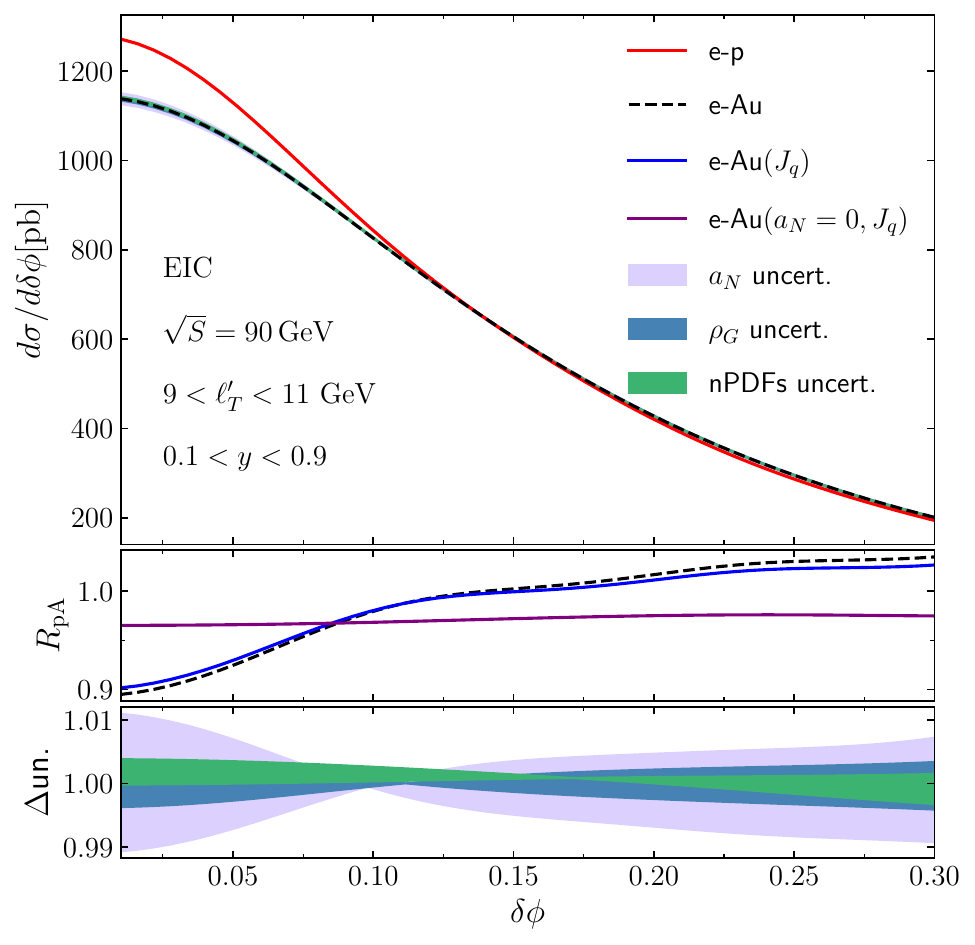}
    \caption{The $\delta\phi$ distribution in electron-proton and electron-gold collisions. The red line is the e-p prediction while we present three different results for e-Au collisions discussed in the text. Besides, we also present the theoretical uncertainties associated with collinear nPDF (green bands), $\rho_G$ (blue bands), and $a_N$ (purple bands), where we vary the $\rho_G$ parameter by a factor of $1-3$, and choose $a_N=0.016\pm0.003$ GeV$^2$.}
    \label{fig:ratio}
\end{figure}

In Fig.~\ref{fig:ratio}, we present the distribution of the azimuthal angle difference, $ \delta\phi $, for e-p collisions depicted in red, alongside three distinct predictions for electron-gold (e-Au) collisions. These predictions are formulated within an energy configuration of $ \sqrt{S} = 90 $ GeV, where we adjust the energy of the incoming nuclear beam by a factor of $ \sqrt{Z/A} $ as suggested in Ref. \cite{Accardi:2012qut}. Our calculations span the region $ 9 \, \text{GeV} < \ell_T' < 11 \, \text{GeV} $ and $ 0.1 < y < 0.9 $. The black dashed curve represents the contributions inclusive of all medium effects. The blue solid curve illustrates the scenario where the jet function is assumed to be identical to that in a vacuum. Lastly, the purple curve corresponds to the case where $ a_N = 0 $ and the vacuum jet function is employed. Centrally positioned in the plot are the ratios of the cross-sections. We observe suppression at low $ \delta \phi $ and enhancement at high $ \delta \phi $, which are indicative of transverse momentum broadening. At extremely large $ \delta \phi $, minor oscillations appear, likely due to numerical instabilities when transitioning out of the TMD region. By contrasting the three curves, it becomes evident that the most significant broadening effects arise from the $ a_N $ contribution, while the jet broadening plays a comparatively minor role. At the bottom of the plot, we outline the theoretical uncertainties, encapsulated by collinear nPDF (green bands), $ \rho_G $ (blue bands), and $ a_N $ (purple bands). It is manifest that the predominant uncertainties currently originate from the nTMD PDFs. Our numerical findings suggest that, given the current parameter values for $ a_N $ and $ \chi $, the process is primarily sensitive to the initial state's broadening effects, thereby serving as a clean probe of nTMD PDF. While a comprehensive comparison with event generators like eHIJING \cite{Ke:2023xeo} is beyond the scope of this paper, it remains an avenue for future investigation.

\section{Summary}
\label{sec:Sum}

In this study, we have focused on the lepton-jet correlation in both e-p and e-A collisions. Utilizing SCET, we derived a factorization theorem tailored for back-to-back lepton-jet configurations in these collisions. The central innovation lies in our adoption of a recoil-free jet axis, which replaces the standard jet axis used in previous works. This change facilitates a more straightforward factorization and resummation formula and enables us to perform all-order resummations at high accuracy for lepton-jet azimuthal distributions.

We have shown that our method elevates the level of perturbative accuracy while mitigating complexities associated with redundant soft factors and the resummation of NGLs present in standard jet axis analyses. Our approach acts as a complementary counterpart to SIDIS investigations, providing an alternative pathway for conducting global analyses of TMD PDFs. This comes with distinct advantages, particularly in terms of streamlined factorization formula and enhanced perturbative resummation accuracy. In the context of e-A collisions, we discussed the utility of our approach in disentangling intrinsic non-perturbative contributions from nTMDs and dynamical medium effects in nuclear environments. Specifically, we argued that the recoil-free jet axis is particularly useful in these collisions, as it can suppress background from underlying events.

Our work sets the groundwork for future experiments at the EIC, aiming to provide a more comprehensive understanding of the inner quark-gluon structure of nucleons and nuclei. Furthermore, our findings hold significant implications for the development of new probes in e-A collisions, offering a robust framework for measuring nTMDs.

\acknowledgments
The authors thank Mei-Sen Gao, Zhong-bo Kang, Ivan Vitev, Wouter Waalewijn, and Yang-Li Zeng for helpful discussion. S.F. and D.Y.S. are supported by the National Science Foundations of China under Grant No.~12275052 and No.~12147101 and the Shanghai Natural Science Foundation under Grant No.~21ZR1406100. W.K. is supported by the U.S. Department of Energy, Office of Science, Office of Nuclear Physics through  Contract No. 89233218CNA000001 and by the Laboratory Directed Research and Development Program at LANL. J.T. is supported by the Department of Energy at LANL through the LANL/LDRD Program under project number~20220715PRD1.

\appendix

\section{Anomalous dimensions up to NNLL}\label{sec:anom}
The cusp and quark non-cusp anomalous dimensions have following expansion in the strong coupling constant
\begin{align}
    \gamma_i(\alpha_s) = \sum_{n=0}^\infty\gamma^i_n\left(\frac{\alpha_s}{4\pi}\right)^{n+1}, \quad{\rm with} \quad i={\rm cusp},~ q,~B,~\mathcal{J},s,r,
\end{align}
and at NNLL the coefficients are given by~\cite{Korchemsky:1987wg, Moch:2004pa,Moch:2005id,Moch:2005tm,Idilbi:2005ni,Idilbi:2006dg,Becher:2006mr}
\begin{align}
\label{eq:c1}
  \gamma_{0}^{\rm cusp} = & \, 4 \,,
  \\
  \gamma_{1}^{\rm cusp} = & \, C_A  \left(\frac{268}{9}-8 \zeta_2\right)-\frac{40  n_f}{9} \nn \,,
  \\
  \gamma_{2}^{\rm cusp} = & \, C_A^2  \left(-\frac{1072 \zeta_2}{9}+\frac{88
      \zeta_3}{3}+88 \zeta_4+\frac{490}{3}\right)  +C_A  n_f \left(\frac{160 \zeta_2}{9}-\frac{112
    \zeta_3}{3}-\frac{836}{27}\right) \nn \\
    & + C_F n_f \left(32 \zeta_3-\frac{110}{3}\right)-\frac{16
                 n_f^2}{27} \,,
\nn
\end{align}
and
\begin{align}
  \gamma^{q}_0 & =   -3\, C_F \,, \\
  \gamma^{q}_1 & =   C_A C_F \left(-11 \zeta_2+26
    \zeta_3-\frac{961}{54}\right) +C_F^2 \left(12 \zeta_2-24 \zeta_3-\frac{3}{2}\right)+C_F n_f \left(2 \zeta_2+\frac{65}{27}\right) \,. \nn \\
 \gamma_0^{B/\mathcal{J}}&=  6 C_F \,, \nn \\
 \gamma_1^{B/\mathcal{J}}&=  C_F^2\left(3-4 \pi^2+48 \zeta_3\right)+C_F C_A\left(\frac{17}{3}+\frac{44 \pi^2}{9}-24 \zeta_3\right)+C_F T_F n_f\left(-\frac{4}{3}-\frac{16 \pi^2}{9}\right). \nn
\end{align}
Besides, the soft anomalous dimensions are
\begin{align}
    \gamma_{0}^s = & \, 0 \,, \\
    \gamma_{1}^s = & \, C_A \left(\frac{22 \zeta_2}{3}+28 \zeta_3-\frac{808}{27}\right)+ n_f \left(\frac{112}{27}-\frac{4 \zeta_2}{3}\right) \,. \nn
\end{align}
Lastly, the rapidity anomalous dimensions are given by
\begin{align}
\gamma_0^r =& \, \gamma_0^s \,, \nn
\\
\gamma_1^r = & \, \gamma_1^s - 2 \zeta_2 \beta_0  \,,
\end{align}
with $\beta_0=11/3\,C_A-4/3\,T_F n_f$.

\section{One loop expressions}

$H$ and $S$ are the hard and soft functions, which are given at one loop to be
\begin{align}\label{eq:loop-hard}
    H\left(Q,\mu\right) = 1+\frac{\alpha_s C_F}{2\pi}\left[-L_Q^2-3L_Q-8+\frac{\pi^2}{6}\right]\,,
\end{align}
\begin{align}\label{eq:loop-soft}
    S (b,n\cdot n_J,\mu,\nu) = 1 +\frac{\alpha_s C_F}{4\pi}\left[-2 L_b^2+4 L_b \ln\left(\frac{\mu^2}{\nu^2}\frac{2}{n \cdot n_J}\right)-\frac{\pi^2}{3}\right]\,, 
\end{align}
\begin{align}\label{eq:loop-softnn}
    S_{n \bar{n}} (b,\mu,\nu) = 1  +\frac{\alpha_s C_F}{4\pi}\left[-2 L_b^2+4 L_b \ln\left(\frac{\mu^2}{\nu^2}\right)-\frac{\pi^2}{3}\right]\,,
\end{align}
with $L_Q = \ln\left(\mu^2/Q^2\right)$, $L_b = \ln\left(\mu^2/\mu_b^2\right)$. The anomalous dimensions for the hard and soft functions for this process are given in Eq. \eqref{eq:anom-soft} while the anomalous dimension of the back-to-back soft function is given by
\begin{align}\label{eq:anom-soft-nn}
    \gamma_\mu^{S_{n\bar{n}}}\left(\mu,\frac{\mu}{\nu}\right) & = \gamma_\mu^{S}\left(\mu,\frac{\mu}{\nu}\right)\Big|_{n\cdot n_J \rightarrow 2}\,,\\
    \gamma_\nu^{S_{n\bar{n}}}\left(b,\mu\right) & = \gamma_\nu^{S}\left(b,\mu\right)\Big|_{n\cdot n_J \rightarrow 2}\,. \nn
\end{align}
Up to one loop, the expressions for the matching coefficients and the jet function are
\begin{align}\label{eq:loop-coll-sub-qq}
    C_{q\leftarrow q}(x,b,\mu,\zeta_f) &= \delta\left(1-x\right)\nn  \\
    &\hspace{-1cm}+ \frac{\alpha_s C_F}{4\pi} \left[-2 L_b P^{(0)}_{qq}(x)+2(1-x)  +\,\delta\left(1-x\right)\left(-L_b^2+2L_b L_{\zeta_f}+3L_b-\frac{\pi^2}{6}\right)\right]\,, \nn\\
     C_{q\leftarrow g}(x,b,\mu,\zeta_f) &= \frac{\alpha_s T_F}{4\pi}\left[-2L_b P_{gq}^{(0)}(x)+4 x(1-x)\right]\,,
\end{align}
with the one-loop splitting functions
\begin{align}
     P_{qq}^{(0)}(x)&=\frac{1+x^2}{(1-x)_+}+\frac{3}{2}\delta\left(1-x\right)\,,\nn\\
     P_{qg}^{(0)}(x)&=x^2+(1-x)^2\,,
\end{align}
and $L_{\zeta_f} = \ln\left(\mu^2/\zeta_f\right)$. While the expression for the jet function is
\begin{align}\label{eq:loop-jet-sub}
     J_q\left(b,\mu,\zeta_J\right) = 1 +\frac{\alpha_s C_F}{4\pi}\left[-L_b^2 + 3 L_b -\frac{5\pi^2}{6}+7-6\,\ln2+2 L_b L_{\zeta_J}\right]\,. 
\end{align}
with $L_{\zeta_J} = \ln\left(\mu^2/\zeta_J\right)$.

\bibliography{refs.bib}

\bibliographystyle{JHEP}

\end{document}